\documentclass[twocolumn,showpacs,preprintnumbers,amsmath,amssymb]{revtex4}
\usepackage{graphicx}
\usepackage{dcolumn}
\usepackage{bm}
\begin{document}
\title{Antiferromagnetic phase of the Kondo-insulator}
\author{R. Eder$^{1}$ and P. Wr\'obel$^{2}$}
\affiliation{$^{1}$ Karlsruhe Institut of Technology, 
Institut f\"ur Festk\"orperphysik, 76021
Karlsruhe, Germany \\
$^{2}$ Institute for Low Temperature and Structure Research, P.O. Box 1410,
50-950 Wroc{\l}aw 2, Poland}
\date{\today}
\abstract{
We discuss the quasiparticle band structure of the antiferromagnetic phase
of the planar Kondo lattice model with a half-filled conduction band, the 
so-called Kondo insulator. The band structure is obtained by bond fermion
technique and good agreement is obtained with the single particle spectral
function obtained by Dynamical Cluster Approximation. In particular,
various changes of the band structure with $J/t$ observed in the numerical
spectra are reproduced qualitatively by the calculation.
In the absence of Fermi surface nesting we find a semimetallic phase for 
sufficiently small exchange constant and possible experimental consequences 
are discussed.}
}
\maketitle
\section{Introduction}
Magnetic ordering transitions in Heavy Fermion compounds continue to be a 
subject of considerable interest in solid state physics. In addition to the 
paramagnetic low-temperature phase with the characteristic heavy bands and a 
Fermi surface volume corresponding to itinerant $f$ 
electrons\cite{Stewart,kondoinsulators}, many of these compounds also
have several antiferromagnetic phases with differ in the ordering wave 
vector of the magnetic moments and/or the Fermi surface volume i.e. including 
the $f$-electrons or not. Often these transitions can be tuned by external 
parameters resulting in quantum critical points, non-Fermi liquid
behaviour and superconducting 
domes\cite{StewartII,loenireview,Steglichreview}.\\
The appropriate model to study Heavy Fermions is the Kondo lattice
model (KLM) which in its simplest form can be written as
\begin{eqnarray}
H&=&\sum_{{\bm k},\sigma}\;\epsilon_{\bm k}^{}\;c_{{\bm k},\sigma}^\dagger 
c_{{\bm k},\sigma}^{} + J\sum_j \;{\bm S}_j\cdot{\bm \sigma}_j.
\label{kondola}
\end{eqnarray}
The model is defined on a lattice (in the present work: a planar s.c. 
lattice) of $N$ unit cells, whereby each unit cell $j$ contains one conduction 
band 
(or $c$) orbital and one localized (or $f$) orbital, the operators
$c_{j,\sigma}^\dagger$ and $f_{j,\sigma}^\dagger$
create an electron with z-spin $\sigma$ in these orbitals.
Moreover, ${\bm \sigma}_j= \frac{1}{2}\;c_{j,\sigma}^\dagger \;
{\bm \tau}_{\sigma \sigma'}\; c_{j,\sigma'}^{}$, 
with ${\bm \tau}$ the vector of Pauli matrices, is the spin
operator for conduction electrons whereas ${\bm S}_j$ denotes the
spin operator for $f$ electrons, defined in an analogous way. An important
feature of the model is the constraint to have precisely one 
electron per $f$-orbital:
\begin{eqnarray}
\sum_\sigma\;f_{j,\sigma}^\dagger f_{j,\sigma}^{} &=& 1,
\label{fconst}
\end{eqnarray}
which must hold separately for each unit cell $j$. The number $N_c$
of conduction electrons is variable, we denote their density/unit cell
as $n_c$, the total electron density then is $n_e=1+n_c$.
Finally
\[
\epsilon_{\bm k}=-2t\left(\;\cos(k_x)+\cos(k_y)\;\right) +4t_1\;
\cos(k_x)\;\cos(k_y)
\]
is the dispersion relation of the conduction band, parameterized by
a nearest-neighbor hopping integral $-t$ and
2$^{nd}$ nearest neighbor hopping integral $t_1$. \\
The KLM is discussed mainly in mean-field (or saddle-point)
approximation, whereby the exchange term which is quartic in electron
operators is subject to mean-field 
factorization\cite{YoshimoriSakurai,LacroixCyrot,Lacroix,AuerbachLevin,Burdinetal,ZhangYu,Lavagna,Senthil,Global,ZhangSuLu,Nilsson},
or by Gutzwiller projection of a suitable trial wave function\cite{Fazekas}.
There has also been a number of numerical studies of the model,
via density matrix renormalization group 
calculations\cite{yuwhite,MC1,MC2,Mutou,Smerat},
quantum Monte-Carlo\cite{Assaad}, series expansion\cite{series,seriesexp}
variational Monte-Carlo (VMC)\cite{WatanabeOgata,Asadzadeh,Kubo} or the
Dynamical Cluster Approximation (DCA)\cite{MartinAssaad,MartinBerxAssaad}.
It is widely believed\cite{Doniach} that magnetic ordering transitions 
in the Heavy Fermion compounds result from a 
competition between the Kondo effect\cite{RG} which favours the paramagnetic
phase and the RKKY interaction\cite{RKKY} between $f$-moments which favours
finite magnetic moments. It should be noted that both, the Kondo effect
and the RKKY-interaction, are adequately described by
the Hamiltonian (\ref{kondola}), so that no additional Heisenberg exchange
between the $f$-electron spins need to be included.
Many studies have aimed at clarifying the nature of these 
transitions\cite{KDB1,KDB2,KDB3,KDB4,KDB5,KDB6,KDB7,KDB8,KDB9,KDB10,KDB11} 
but a consensus regarding the nature of these
has not yet been achieved. One controversial question is whether
the heavy quasiparticles persist at the magnetic transition, so that
this may be viewed as the heavy bands undergoing a conventional
spin density wave transition, or whether the magnetic ordering
suppresses the Kondo effect, so that the heavy bands disappear alltogether.\\
It is the purpose of the present manuscript to discuss the band structure of 
the antiferromagnetic phase for $n_e=2$ in the framework of 
bond fermion theory\cite{Oana,JureckaBrenig,afbf}. 
It was shown recently\cite{afbf} that bond fermion theory reproduces the 
phase diagram in the $(J/t,n_c)$ plane obtained by 
VMC\cite{WatanabeOgata,Asadzadeh,Kubo} or Dynamical Mean Field 
Theory (DMFT)\cite{PetersKawakami} for the planar KLM quite well.
More precisely, it was found that on one hand bond fermion theory gives a 
too large value for $J_{c,1}/t$, defined as the value of $J/t$ where 
antiferromganetic order sets in at half-filling.
On the other hand, if the phase diagram is plotted as a function
of $(J/J_{c,1},n_c)$ rather than $(J/t,n_c)$, so that the error in 
$J_{c,1}$ cancels out to some degree, it agrees quite well
with the one obtained by the numerical 
methods, see Figure 7 in Ref. \cite{afbf}.
This is remarkable in that the phase diagram of the planar KLM is quite
intricate, including the paramagnetic and two antiferromagnetic phases
with different Fermi surface topology divided by a Lifshitz transition.
Moreover, even for numerical methods it appears to be difficult to 
correctly reproduce $J_{c,1}/t$: for $t_1=0$ the exact value is 
$J_{c,1}/t=1.45$\cite{Assaad}, VMC finds $J_{c,1}/t=1.7$\cite{WatanabeOgata},
DMFT finds $J_{c,1}/t=2.2$\cite{PetersKawakami} whereas DCA gives 
$J_{c,1}/t=2.1$\cite{MartinBerxAssaad}. 
The results of Ref. \cite{afbf} thus show that regarding the phase diagram 
bond Fermion theory gives a `rescaled version' of the actual physics.
In the present manuscript we focus on details of the single particle
spectrum - i.e. the correlated band structure - for $n_e=2$ and compare in 
detail to recent DCA calculations
by Martin {\em et al.}\cite{MartinBerxAssaad}. Since the DCA calculation
finds antiferromgnetic order at $n_e=2$ we disregard incommensurate
or stripe-like order which may occur for metallic densities\cite{stripes}.
It will be seen that
bond Fermion theory reproduces the single particle spectral density
quite well and even subtle changes of the quasiparticle bands
with $J/t$ are reproduced, provided one rescales $J$ and gap energies 
by $J_{c,1}$.
\section{Formalism}
We study the KLM for the case $n_e=2$, that means a half-filled
conduction band. For $t_1=0$ - which means a $(\pi,\pi)$-nested Fermi surface
for the decoupled conduction electrons - it is known that antiferromagnetic
ordering occurs for $J/t\le J_{c,1}/t=1.45$\cite{Assaad} and the
value of $J_{c,1}$ may be expected to be smaller for finite $t_1$.
Bond fermion theory is similar in spirit to the
bond boson theory for spin systems\cite{SachdevBhatt,Gopalan} and 
amounts to mapping a subset
of states of the true KLM to a fictitious Hilbert space
of bond fermions. More precisely, we first define the following operators
and states:
\begin{eqnarray}
s_j^\dagger&=&\frac{1}{\sqrt{2}}\;(c_{j,\uparrow}^\dagger f_{j,\downarrow}^\dagger - c_{j,\downarrow}^\dagger f_{j,\uparrow}^\dagger),\nonumber\\
t_{j,z}^\dagger&=&\frac{1}{\sqrt{2}}\;
(c_{j,\uparrow}^\dagger f_{j,\downarrow}^\dagger 
+ c_{j,\downarrow}^\dagger f_{j,\uparrow}^\dagger),\\
\tilde{s}_j^\dagger &=&\cos(\Theta)\; s_j^\dagger 
+ e^{i{\bm Q}\cdot {\bm R}_j}\sin(\Theta)\; t_{j,z}^\dagger\nonumber \\
|\Psi_0\rangle &=&\prod_{j=1}^N\;\tilde{s}_j^\dagger\; |0\rangle,
\label{vacuum_1}
\end{eqnarray}
with ${\bm Q}=(\pi,\pi)$.
The operators $s_j^\dagger$ and $t_{j,z}^\dagger$ create states of one conduction 
and one $f$-electron in unit cell $j$, whereby the spins of the electrons are
coupled to a singlet or triplet.
The superposition of these states, created by $\tilde{s}_j^\dagger$,
has an energy of 
$-\tilde{e}_0 = -\frac{3J}{4}\cos^2(\Theta) + \frac{J}{4}\sin^2(\Theta)$
and a nonvanishing expectation value
$\langle {\bm S}_i\rangle =-e^{i{\bm Q}\cdot {\bm R}_j} \sin(2\Theta)/2\;{\bm e}_z$.
Accordingly, $|\Psi_0\rangle$
is an antiferromagnetic state (for $\Theta\ne 0$) with 
two electrons per unit cell and the expectation value of the energy
is $-N\tilde{e}_0$. It may be viewed as a condensate of triplets
into momentum ${\bm Q}$\cite{SachdevBhatt} on a background of singlets.\\
Let us now assume that starting from $|\Psi_0\rangle$ the hopping term for
the $c$-electrons is switched on.
Under the action of the hopping term, $c$-electrons are
transferred between unit cells so that there will also be cells containing
either a single or three electrons. In bond fermion theory cells
with an odd number of electrons are interpreted as occupied by 
Fermions. More precisely, a cell $j$ in the state $f_{j,\sigma}^\dagger|0\rangle$
is considered occupied by a hole-like Fermion, created
by $a_{j,\sigma}^\dagger$, in the bond fermion Hilbert space,
whereas if the cell is in the state  
$c_{j,\uparrow}^\dagger c_{j,\downarrow}^\dagger f_{j,\sigma}^\dagger|0\rangle$
it is considered occupied by an electron-like Fermion, created
by $b_{j,\sigma}^\dagger$. 
Denoting the set of cells occupied by a single electron (three electrons) by 
$S_a$ ($S_b$) and defining $S_s$ as the
complement of $S_a\cup S_b$ (that means $S_s$ is the set of cells with two 
electrons) the correspondence between the bond fermion states and the states 
of the KLM is
\begin{eqnarray}
\left(\;\prod_{i\in S_a} a_{i,\sigma_i}^\dagger\;\right)
\left(\;\prod_{j\in S_b} b_{j,\sigma_j}^\dagger\;\right)|0\rangle \rightarrow
\;\;\;\;\;\;\;\;\;\;\;\;\;\;\;\;\;\;\;\;\;\;\;\;\;\;\;\;\;\;\;\;\nonumber \\
\left(\;\prod_{i\in S_a} f_{i,\sigma_i}^\dagger\;\right)
\left(\;\prod_{j\in S_b} c_{j,\uparrow}^\dagger c_{j,\downarrow}^\dagger f_{j,\sigma_j}^\dagger\;\right)
\left(\;\prod_{n\in S_s}\tilde{s}_n^\dagger\;\right)\;|0\rangle. \nonumber \\
\label{trans}
\end{eqnarray}
The Hamiltonian $H_{BF}$ (or any other operator) 
for the bond fermions is now derived by demanding
that its matrix elements between the states on the left hand side
of (\ref{trans}) are equal to those of the true KLM Hamiltonian 
(\ref{kondola}) between the corresponding states on the right hand side of 
(\ref{trans}).  In particular, the electron creation operators, from which 
many other operators can be constructed, become
\begin{eqnarray}
c_{j,\uparrow}^\dagger&=&
\frac{1}{\sqrt{2}}\left(\;\;\,\zeta_{j}^{(+)} a_{j,\downarrow}^{} 
- \zeta_{j}^{(-)} b_{j,\uparrow}^\dagger\;\right),
\nonumber \\
c_{j,\downarrow}^\dagger&=&
\frac{1}{\sqrt{2}}\left(-\zeta_{j}^{(-)} a_{j,\uparrow}^{} - \zeta_{j}^{(+)} b_{j,\downarrow}^\dagger\;\right),
\nonumber \\
\zeta_{j}^{(\pm)} &=&\cos(\Theta)\pm e^{i{\bm Q}\cdot {\bm R}_j}\sin(\Theta),
\label{resolution}
\end{eqnarray}
whereas the exchange term in (\ref{kondola}) takes the form
\begin{eqnarray}
H_J &=&
\tilde{e}_0 \sum_j \left(\;b_{j,\sigma}^\dagger b_{j,\sigma}^{}
+ a_{j,\sigma}^\dagger a_{j,\sigma}^{}\;\right) - N\tilde{e}_0.
\label{fermex}
\end{eqnarray}
It is obvious that in order for (\ref{trans}) to make sense,
$S_a$ and $S_b$ have to be disjunct. This is equivalent to the constraint
on the bond fermions that no two of them occupy the same site,
which in turn is equivalent to an infinitely strong repulsion between them.
It is known\cite{Nozieres} that in the limit 
$J/t\rightarrow \infty$  and $N_c \ne N$ the KLM is equivalent to a 
$U/t=\infty$ Hubbard model for $|N-N_c|$ `bachelor spins'. The  
$a^\dagger$-Fermions ($b^\dagger$-Fermions) then obviously correspond to these 
bachelor spins for $N_c<N$ ($N_c>N$). For finite $J/t$ the two types of 
Fermions coexist, but are subject to an infinitely strong mutual repulsion.
However, as shown in  Ref. \cite{afbf}, the density of the $a^\dagger$ and 
$b^\dagger$ Fermions is quite small over large regions of parameter space so 
that the constraint can be relaxed to good approximation. In principle even 
such an infinitely strong repulsion in a low density Fermi gas
can be treated using known methods 
from field theory\cite{FW}. In the case of bond boson theory for spin systems 
this was in fact carried out by Kotov {\em et al.}\cite{Kotov} and 
Shevchenko {\em et al.}\cite{Shevchenko}. 
Since there are several species of Fermions this would be more complicated 
for the bond fermions and in the 
following we simply relax the constraint and treat the fermions as 
noninteracting. It will be seen that even in this simplest approximation
the results are not too bad.
Equation (\ref{trans}) also shows the main advantage of bond fermion 
theory: all basis states fulfil the constraint (\ref{fconst})
exactly, so that this is `hard wired' in bond fermion theory.
On the other hand, it is obvious from the above that bond Fermion theory is
by nature a strong coupling theory so that one cannot expect it to
reproduce the energy scale of the single impurity Kondo temperature 
$T_K=W\;e^{-1/\rho J}$, which emerges in the limit of small $J/t$
($W$ and $\rho$ are the bandwidth and density of states of the
conduction band). In fact, for
$J/t\rightarrow 0$ the density of the fermions increases strongly so 
that relaxing the constraint cannot be expected to be a meaningful 
approximation anymore.\\
The derivation of $H_{BF}$ is given
in Ref. \cite{afbf} and since the formulas are somewhat lengthy we do not 
reproduce them here. We note, however, that $H_{BF}$ is quadratic in
fermion operators - which is possible because the exchange term
$H_J$ becomes a quadratic from in `bond fermion language', see
Eq. (\ref{fermex}) - so once we relax the constraint of no double occupancy it 
can be readily diagonalized by a unitary transformation. Since there are two 
types of Fermions/site  and two sublattices there are four bands, denoted
by $E_{\nu,\bf{k}}$.
Knowing the band structure one can calculate the ground state energy $E_0$ as 
a function of the as yet undetermined angle $\Theta$ in (\ref{vacuum_1}). In 
the last step the angle $\Theta$, which controls the degree of admixture of the
triplet and thus the magnitude of the ordered moment, is fixed
by minimizing $E_0$. \\
\begin{figure}
\resizebox{0.40\textwidth}{!}{%
\includegraphics{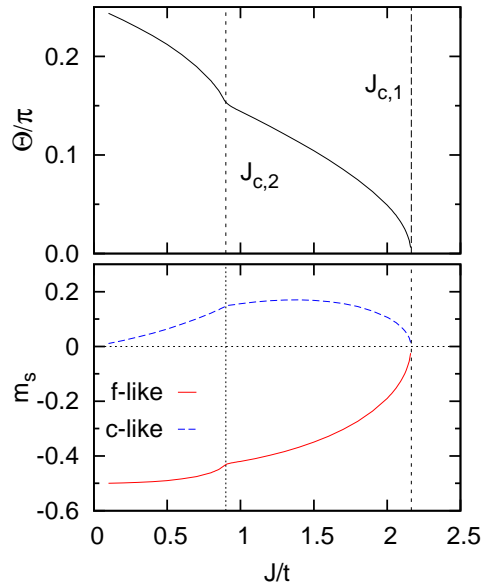}
}
\caption{Top: Optimum angle $\Theta$ as a function of $J/t$,
$n_e=2$, $t_1/t=0.3$. Bottom: Ordered moment versus $J/t$.}
\label{fig1}
\end{figure}
\section{Results}
Figure \ref{fig1} shows the angle $\Theta$ which minimizes 
$E_0$ as a function of $J/t$ for $t_1/t=0.3$. At approximately
$J/t=2.165$ the optimum value of  $\Theta$ starts to deviate from zero,
indicating a second order transition to the antiferromagnetic phase.
It should be noted that no direct Heisenberg exchange between $f$-electrons is 
included in the Hamiltonian, rather this transition is caused solely by the
`implicit' interaction between $f$-spins mediated by the conduction electrons.
At $J/t=0.9$ there is an anomaly - i.e. a pronounced
upward bend in the curve. This anomaly is absent in the case
$t_1/t=0$ - see Figure 1 of  Ref. \cite{afbf}.
The Figure also shows the $f$-like ordered moment
\begin{eqnarray*}
m_{s,f}&=&\frac{1}{N}\;\sum_j\;e^{i{\bm Q}\cdot {\bm R}_j} \langle S_{j,z}\rangle,
\end{eqnarray*}
and an analogous definition for the $c$-like moment.
The ordered moments deviate from zero at $J_{c,1}$
and also show the anomaly at $J/t=0.9$.
The behavior of the $f$-like ordered moment for $J/t\rightarrow 0$ is somewhat 
surprising in that its magnitude
approaches the saturation value of $1/2$. It should be 
noted, however, that exactly the same behavior is seen in the Quantum Monte 
Carlo data for $t_1/t=0$ in Ref. \cite{Assaad} which up to statistical errors
are exact results. This highlights the fact that $J/t=0$ is a singular point 
of the model.
\begin{figure}
\resizebox{\columnwidth}{!}{%
\includegraphics{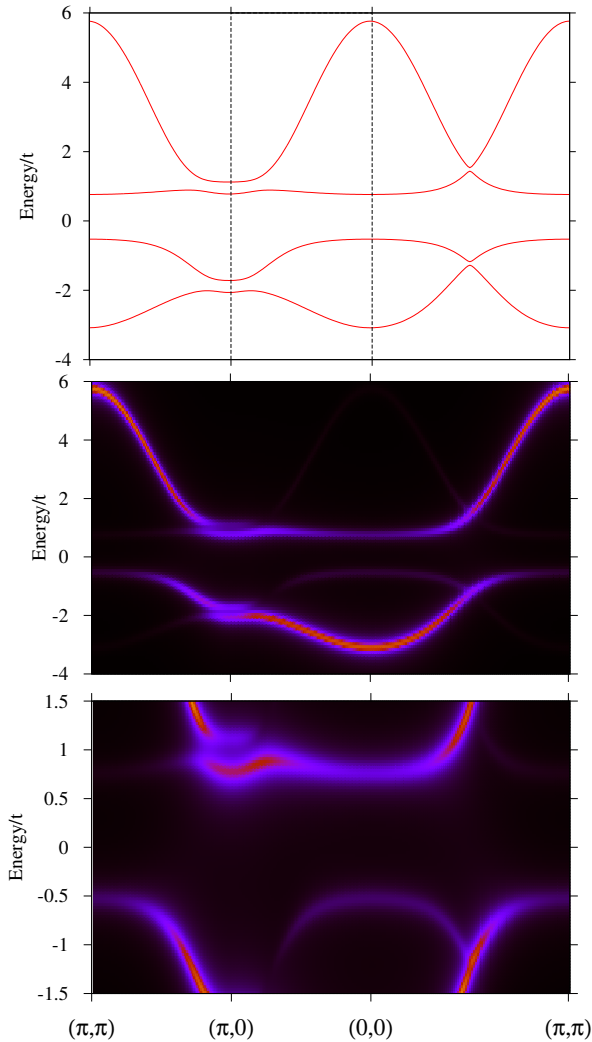}
}
\caption{Band structure $E_{\nu,\bf{k}}$ (topmost panel)
and spectral density $\rho({\bm k},E)$
(lower two panels) for the antiferromagnetic
phase of the Kondo insulator, $J/t=1.9$, $t_1/t=0.3$. The chemical
potential is the zero of energy.}
\label{fig2}
\end{figure}
In the following we denote the value of $J$ 
where the anomaly occurs by $J_{c,2}$.
In Ref. \cite{afbf} it was found that for the case $t_1=0$
bond fermion theory predicts the value $J_{c,1}/t=2.3$. As already
mentioned this is too large compared to the exact value 
$J_{c,1}/t=1.45$\cite{Assaad}, but when $J$ is measured 
in units of $J_{c,1}$ so that the error in $J_{c,1}/t$ cancels out to some 
degree, the phase diagram from bond fermion theory is in good agreement 
with numerical results. Basically the same will be seen to hold true
for the band structure.\\
\begin{figure}
\resizebox{0.80\columnwidth}{!}{%
\includegraphics{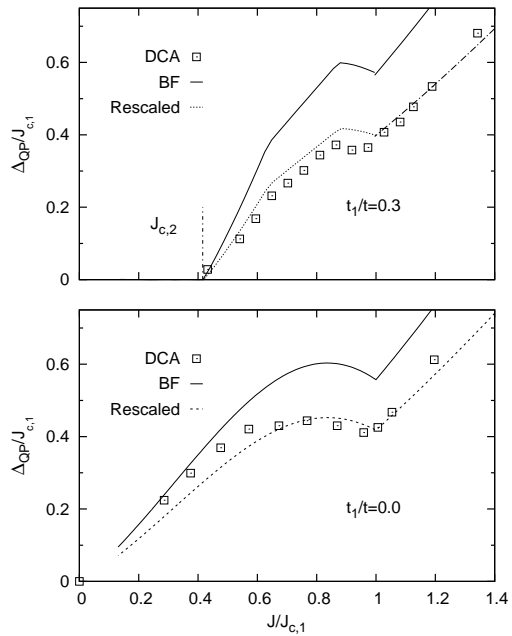}
}
\caption{Quasiparticle gap $\Delta_{QP}$ for $t_1/t=0.3$ (top)
and $t_1/t=0.0$ (bottom) from bond fermion theory
(BF) and rescaled bond fermion theory as described in the text
(Rescaled) compared to DCA.}
\label{fig3}
\end{figure}
To begin with, Figure \ref{fig2} shows the band structure $E_{\nu,\bf{k}}$ 
and $c$-like spectral density
\[
\rho({\bm k},E)=-\frac{1}{\pi}\;Im\;G({\bm k},E+i0^+),
\]
for the antiferromagnetic phase. Here $G({\bm k},E)$ is the $c$-electron
Green's function which is readily obtained from the eigenvalues und eigenvectors
of $H_{BF}$ and the representation (\ref{resolution}).
Whereas the band structure shows
antiferromagnetic (AF) symmetry this is not at all the case for the
spectral density. The individual bands have a strongly ${\bm k}$-dependent
spectral weight and with the exception of the heavy bands
forming the gap around the chemical potential
the AF-umklapps have hardly any spectral weight.
Along $(0,0)\rightarrow (\pi,\pi)$ in particular this creates the impression 
as if the strongly dispersive $c$-like band branches into two almost
dispersionless bands of low spectral weight. Much the same
can be seen in the spectral function obtained by DCA, see Figure 4 in 
Ref. \cite{MartinBerxAssaad} and in fact the whole spectral density is quite
similar to DCA.\\
The nature of the anomaly in the $\Theta$ v.s. $J/t$ curve
at $J_{c,2}$ seen in Figure \ref{fig1} becomes clearer in
Figure \ref{fig3}. This shows the quasiparticle gap
\[
\Delta_{QP} = E_0(N_e+1)+E_0(N_e-1) -2E_0(N_e),
\]
as a function of $J/J_{c,1}$.
Thereby $E_0(N_e)$ is the ground state energy for $N_e$ electrons.
In a system described by bands of noninteracting quasiparticles
this is the energy gap between the highest occupied and lowest unoccupied 
energy of the band structure $E_{\nu,\bf{k}}$. 
For $J/J_{c,1}> 1$ there is always a finite gap, that means the system is
a paramagnetic insulator. The gap is quite large and to good approximation
linear in $J$ - with no indication of the exponential dependence of $T_K$
on $J$. At $J_{c,1}$, $\Delta_{QP}$ has an upward kink and then decreases 
roughly linearly with 
$J/J_{c,1}$. For $t_1/t=0.3$, $\Delta_{QP}$ approaches zero at 
$J_{c,2}$, whereas for $t_1=0$ it extrapolates to zero only at $J=0$. 
This highlights the importance of Fermi surface nesting for the
decoupled conduction electron band. Nonvanishing $t_1$ gives
a finite dispersion along the antiferromagnetic zone boundary
and hence an anisotropic gap. Figure \ref{fig3} also shows the values of
$\Delta_{QP}$ obtained by Martin {\em et al.} \cite{MartinBerxAssaad}
by DCA. These authors found $J_{c,1}/t=1.85$, considerably smaller than
the value $J_{c,1}/t=2.165$ from bond fermion theory.
When both $J$ and $\Delta_{QP}$ are measured in units
of $J_{c,1}$, however, the DCA results agree qualitatively with the bond fermion 
curve, in particular the ratio $J_{c,2}/J_{c,1}\approx 0.4$ for
$t_1/t=0.3$ is very similar.
When the bond fermion values for $\Delta_{QP}/J_{c,1}$ are in addition 
rescaled by a phenomenological factor of $0.65$ for $t_1/t=0.3$
and $0.8$ for $t_1/t=0$ the agreement becomes 
almost perfect for $J>J_{c,2}$ where $\Delta_{QP}$ is so large that it can 
be resolved in the DCA calculation (the somewhat zigzag shape of the bond
Fermion curve is due to the fact that the momenta where the maximum of the
lower band/minimum of the upper band are located change with $J/t$).
The question then arises as to what is the nature of the
ground state for $J\le J_{c,2}$. Martin {\em et al.} 
argued\cite{MartinBerxAssaad} that the
system still has a nonvanishing gap even in this parameter range,
whereby this gap traces the single impurity Kondo temperature
$T_K$ which rapidly decreases for small $J/t$
and thus can no longer be resolved by a numerical technique
such as DCA for small enough $J/t$. The bond fermion
calculation suggests a different interpretation:
\begin{figure}
\resizebox{0.40\textwidth}{!}{%
\includegraphics{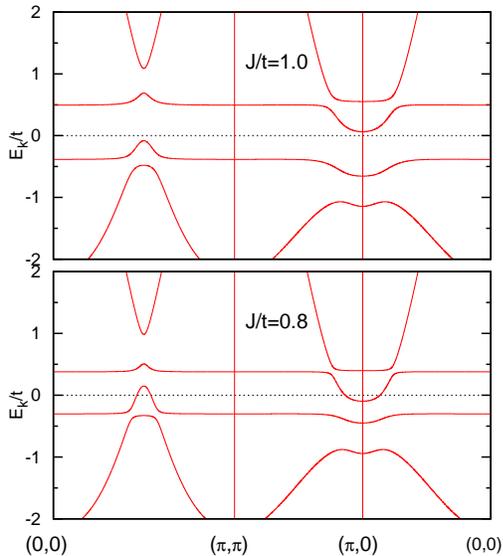}
}
\caption{Band structure around $E_F$ for
$J/t=1.0$ (top) and $J/t=0.8$ (bottom), $t_1/t=0.3$.}
\label{fig4}
\end{figure}
Figure \ref{fig4} shows the band structure for two values of $J/t$,
one above and one below $J_{c,2}$. The Figure shows that
the anomaly corresponds to a transition from an
insulator to a semimetal with an electron pocket around
$(\pi,0)$ and a hole pocket around $(\frac{\pi}{2},\frac{\pi}{2})$.
For $J/t\le 0.9$ the bond fermion calculation therefore
predicts the system to be semimetallic so that $\Delta_{QP}=0$.
On the other hand, the transition to the semimetal might also simply
indicate the breakdown of the bond fermion description due to
its inability to reproduce the energy scale of $T_K$. This
cannot be decided with the information at hand.
Figure \ref{fig5} shows $J_{c,1}$ and $J_{c,2}$ as functions of $t_1/t$.\\
\begin{figure}
\resizebox{0.45\textwidth}{!}{%
\includegraphics{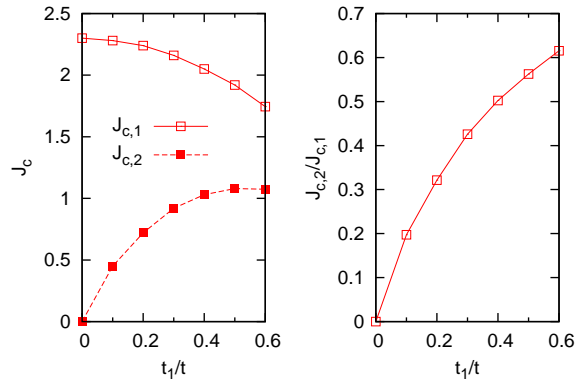}
}
\caption{$J_{c,1}$ and $J_{c,2}$ (left) and the ratio
$J_{c,2}/J_{c,1}$ (right) versus $t_1/t$.}
\label{fig5}
\end{figure}
\begin{figure}
\resizebox{0.30\textwidth}{!}{%
\includegraphics{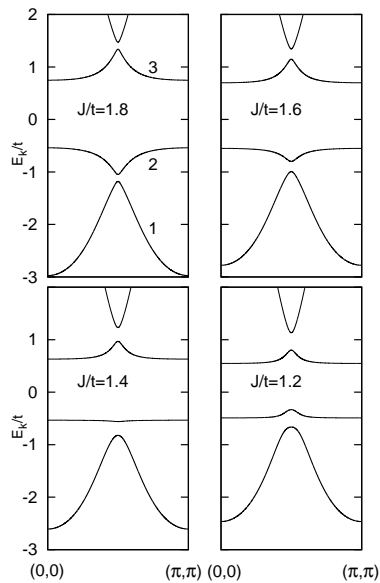}
}
\caption{Band structure around $E_F$ along $(0,0)\rightarrow (\pi,\pi)$
for different values of $J/t$, $t_1/t=0.3$}
\label{fig6}
\end{figure}
We now consider in more detail the evolution of the band structure
with $J/t$ in the range $J>J_{c,2}$.
Figure \ref{fig6} shows the band structure along 
$(0,0)\rightarrow (\pi,\pi)$ for different $J/t$ ($t_1/t=0.3$).
For the relatively large value of $J/t=1.8$ the maximum of the upper
occupied band - labeled 2 in Figure \ref{fig6} - is at $(\pi,\pi)$,
the minimum at $(\frac{\pi}{2},\frac{\pi}{2})$. As $J/t$ decreases,
the minimum at $(\frac{\pi}{2},\frac{\pi}{2})$ becomes 
shallower and at $J/t=1.4$ the band is almost dispersionless.
Decreasing $J/t$ even more `inverts' the dispersion of the band
in that the maximum of the band in question is at  
$(\frac{\pi}{2},\frac{\pi}{2})$ whereas the minimum is at 
$(\pi,\pi)$. Exactly the same has also been observed
by Martin {\em et al.} in their DCA calculation, see
Figure 7 in Ref. \cite{MartinBerxAssaad}.
DCA predicts the value of $J/t$ where  $(\frac{\pi}{2},\frac{\pi}{2})$
changes from being minimum to being maximum to be around
$J/t=1.25=0.68\;J_{c,1}/t$, the bond fermion calculation finds 
$J/t=1.4=0.65\;J_{c,1}/t$. \\
\begin{figure}
\resizebox{0.30\textwidth}{!}{%
\includegraphics{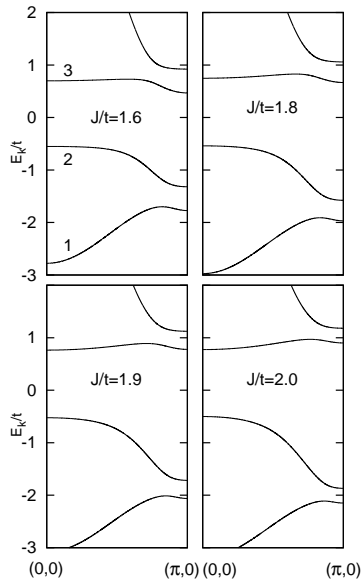}
}
\caption{Band structure around $E_F$ along $(0,0)\rightarrow (\pi,0)$
for different values of $J/t$, $t_1/t=0.3$.}
\label{fig7}
\end{figure}
Another detail of the evolution of the band structure is shown in 
Figure \ref{fig7} which shows the band structure along 
$(0,0)\rightarrow (\pi,0)$ for larger $J/t$. At $J/t=1.6$ the minimum of the 
lower unoccupied band - labeled 3 in Figure \ref{fig7} -
is at $(\pi,0)$ (it is this minimum which crosses below
$E_F$ at $J_{c,2}$). With increasing $J/t$ the difference between
the energies at $(0,0)$ and $(\pi,0)$ becomes smaller
and for $J/t=1.9$ the band has the same energy at these two
momenta. For $J/t=2.0$ the minimum of the band shifts to $(0,0)$.
Again, the same behaviour is seen in the DCA spectra, see Figure 8
of Ref. \cite{MartinBerxAssaad}. DCA finds 
that the minimum shifts at $J/t=1.55=0.84\;J_{c,1}/t$ whereas 
the bond fermion calculation gives $J/t=1.90=0.88\;J_{c,1}/t$.
As was the case for the phase diagram, bond fermion theory appears
to give a `rescaled version of reality':
while it does not reproduce absolute energy scales such as the
correct $J_{c,1}/t$ or the quasiparticle gap accurately,
it reproduces the the band structure and its changes
with $J/J_{c,1}$ quite well.
\section{Summary and Discussion}
In summary we have shown that the bond fermion theory qualitatively 
reproduces a number of results obtained by numerical methods,
in particular the Dynamical Cluster Approximation 
(DCA) for the KLM. The main deficiency is the overestimation of the 
value of $J_{c,1}/t$, where the transition to the antiferromagnetic phase 
occurs. As already mentioned, however, even for numerical methods it is 
difficult to reproduce this value accurately. The single particle 
spectral density in the antiferromagnetic phase is in good agreement with DCA 
and when values of $J$ are measured in units of $J_{c,1}$ the variation of the 
quasiparticle gap with $J/J_{c,1}$ is similar as obtained by DCA, in particular 
the value $J_{c,2}/J_{c,1}$ where the quasiparticle gap (approximately)
closes is reproduced well. Even fine details in the
change of the band structure with $J/J_{c,1}$, such as the shift of band maxima
and minima between different point in the Brillouin zone
are reproduced well by theory.
Interestingly, the shift of the maximum of the topmost occupied band from
$(\pi,\pi)$ to  $(\frac{\pi}{2},\frac{\pi}{2})$
already foreshadows the Lifshitz transition
between the two antiferromagnetic phases for 
$n_e < 2$\cite{WatanabeOgata,Asadzadeh,Kubo,PetersKawakami},
because this is precisely a transition from a pocket around $(\pi,\pi)$ 
to a pocket around $(\frac{\pi}{2},\frac{\pi}{2})$.\\
As already mentioned, bond fermion theory is a strong coupling theory
by nature and cannot reproduce the single impurity energy scale
$T_K$. On the other hand, Quantum Monte Carlo finds
antiferromagnetic ordering in the 2D KLM for the quite large value 
$J/t=1.45$ where the quasiparticle gap varies linearly with
$J/t$ and no exponential dependence on $J$ is observed\cite{Assaad}.
It is unclear if ordering for such a large value of $J/t$
is special for the planar model but this makes
bond Fermion theory useful to discuss antiferromagnetism.\\
The fact that bond fermion theory describes the KLM reasonably
well if $J/t$ is rescaled to lower values can be
understood qualitatively by considering the effects of
the infinitely strong repulsion between the fermions. First, the repulsion 
could lead to a reduction of the the hopping integral, $t_{eff}<t$. Second, 
since presence of
a bond fermion at some site $i$ blocks all hopping processes of
other fermions involving this site, the repulsion should lead to
a loss of kinetic energy per fermion and thus to an increase of the
energy $\tilde{e_0}$ ascribed to a bond fermion in (\ref{fermex}).
Since $\tilde{e_0}\propto J$ this might have a similar effect as
using an effective $J_{eff}>J$ in the bond fermion calculation. 
Both effects would render
$J/t<J_{eff}/t_{eff}$ so if one assumes that the parameters $J$ and $t$ 
in the above calculations actually correspond to the renormalized
$J_{eff}$ and $t_{eff}$ this would explain why the values of
$J/t$ have to be reduced to be consistent with numerics.
In fact, using a version of bond fermion theory which incorporates
the downward renormalization of $t$, Jurecka and Brening found
$J_{c,1}/t=1.505$ for $t_1=0$\cite{JureckaBrenig}, remarkably close to the 
exact value $J_{c,1}/t=1.45$. The way in which the constraint
on the bond fermions must be treated needs additional study.\\
For nonvanishing next-nearest neighbor hopping $t_1$ - that means in the
absence of $(\pi,\pi)$-nesting for the half-filled decoupled conduction band -
bond fermion theory predicts a phase transition to a semimetallic state
for small $J/t$. Assuming that a compression of the material
would increase the ratio $J/t$ - which is plausible because pressure
would tend to increase all hopping integrals and $J$ is proportional to higher 
powers of these - a hypothetical compound which realizes the semimetallic 
phase could be driven to the insulating state by applying pressure. This is 
in contrast to the behaviour of a band insulators with a small gap which
would tend to become semimetallic under pressure.

\end{document}
